\documentstyle{article}
\everymath{\rm }
\everydisplay{\rm }

\begin{document}
\begin{center}
{\LARGE Closed-Flux Solutions to the \\ [.125in] Quantum
Constraints
for\\ [.125in] Plane
Gravity Waves} \\ [.25in]
\large Donald E. Neville \footnote{\large e-mail address:
nev@vm.temple.edu }\\Department of Physics \\Temple University
\\Philadelphia 19122, Pa. \\ [.25in]
July 23, 1996 \\ [.5in]
\end{center}
\newcommand{\E}[2]{\mbox{$\tilde{{\rm E}} ^{#1}_{#2}$}}
\newcommand{\A}[2]{\mbox{${\rm A}^{#1}_{#2}$}}
\newcommand{\Np}{\mbox{${\rm N}'$}}
\newcommand{\Etwo}{\mbox{$^{(2)}\!\tilde{\rm E} $}}
\newcommand{\Etld }{\mbox{$\tilde{\rm E}  $}}
\def \ut#1{\rlap{\lower1ex\hbox{$\sim$}}#1{}}
\newcommand{\phst}{\mbox{$\phi\!*$}}
\newcommand{\bea}{\begin{eqnarray}}
\newcommand{\eea}{\end{eqnarray}}
\newcommand{\be}{\begin{equation}}
\newcommand{\ee}{\end{equation}}
\newcommand{\rta}{\mbox{$\rightarrow$}}
\newcommand{\EQ}[1]{equation~(\ref{eq:#1})}
\Roman{section}
\large
\begin{center}
{\bf Abstract}
\end{center}
The metric for plane gravitational waves is quantized within the
Hamiltonian framework, using a
Dirac constraint quantization and the self-dual field variables
proposed by Ashtekar.  The z axis
(direction of travel of the waves) is taken to be the entire real
line rather than the torus (manifold
coordinatized by (z,t) is RxR rather than $S_1$ x R).
Solutions to the constraints proposed in a previous paper involve
open-ended flux lines running
along the entire z axis, rather
than closed loops of flux; consequently, these solutions are
annihilated by
the Gauss constraint at  interior points
of  the z axis, but not at  the two boundary points.  The solutions
studied in the present paper are based on closed flux loops and
satisfy the Gauss constraint for all z.   \\[.125in]
PACS categories: 04.60.Ds, 04.30.Nk

\section{Introduction}

     This is one of a series of papers discussing quantization of
plane gravitational waves \cite{I,II}.  The emphasis is on the
Ashtekar approach \cite{Ash87,
Alect,Rovlect}; a major goal is to
understand how radiative phenomena are expressed using the Ashtekar
language.  The method of
quantization is canonical.  Due to the high degree of symmetry
(full translational symmetry in
the x,y directions), four of the seven scalar, vector, and Gauss
constraints can be solved and
eliminated from the problem before
quantization.  (The four are: the two vector constraints generating
x,y spatial diffeomorphisms;
and the two
Gauss constraints generating ``internal''   rotations around the X
and Y axes.)   The remaining
three constraints are imposed after quantization, and the key step
is finding a wavefunctional
which is annihilated by these constraints.
      A previous paper (II) wrote down the threee survivng
constraints in the Ashtekar formalism,
quantized them by replacing one half of each (A, \Etld) canonical
pair by a functional derivative,
and wrote down a class of wavefunctionals $\psi$ annihilated by the
the constraints \cite{II}.

     The $\psi$ constructed in II are unusual.  It is usual to have
the $\psi$ depend on the
connection
fields A, in order to
exploit analogies with Yang-Mills theory, whereas these $\psi$
depend on only one connection
field, \A{Z}{z}.  The remaining dependence is on densitized inverse
triad fields \Etld .   Each
$\psi$ is a  string of \Etld\
operators separated by holonomy matrices depending on \A{Z}{z}.
If the holonomies are
visualized in the usual way as lines of flux, then these solutions
are flux lines extending  along
the z axis, with  \Etld\ fields inserted into the line.

     These solutions
satisfy the Gauss constraint at finite z, but not at infinity.
This can also be understood in
visual terms.  The flux line is an open, rather than closed curve,
since the flux line does not loop
back to close on itself.  Since flux exits at infinity, the Gauss
constraint is not satisfied there.

     The present paper constructs solutions which are ``closed
flux''   in nature and  are
annihilated by the Gauss constraint everywhere.  Since the present
paper is for the most part an
extension of II,
I have refrained from repeating many matters of philosophy and
motivation already
discussed in that paper.  I have repeated here some of the more
detailed results of II,  however,
so as to make the calculations
of the present paper self-contained: see the first half of section
2 and the opening paragraphs of
section 3.

     The first half of section 2 reviews the quantization procedure
of paper II and writes out
the three constraints using Ashtekar variables.  The second half of
section 2 constructs a
point-splitting regularization for one of the terms in the scalar
constraint.    (In II it was not
necessary
to regulate this term, because it vanishes identically when acting
on the solutions of II.)   Section
3 presents the ansatz for $\psi$ and shows that the three
constraints annihilate the ansatz.
Section 4 presents conclusions and directions for further research.

           My notation is typical of papers based upon the
Hamiltonian
approach with concomitant 3+1 splitup.  Upper case indices A,
B, $\ldots $,I, J, K, $\ldots$ denote local Lorentz indices
("internal" SU(2) indices) ranging over X, Y, Z only.  Lower case
indices a, b, $\ldots $, i, j, $\ldots $ are also three-
dimensional and denote global coordinates on the three-manifold.
Occasionally the formula will contain a field with a superscript
(4), in which case the local Lorentz indices range over X, Y, Z,
T and the global indices are similarly four-dimensional; or a
(2), in which case the local indices range over X, Y (and global
indices over x, y) only.  The (2) and (4) are also used in
conjunction with determinants; e.\ g., g is the usual 3x3
spatial determinant, while \Etwo\ denotes the determinant of
the 2x2 X, Y subblock of the densitized inverse triad matrix
\E{a}{A}. I use Levi-
Civita symbols of various dimensions: $\epsilon _{TXYZ} =
\epsilon _{XYZ} = \epsilon _{XY} = +1$.  The Minkowski metric
convention is
$\eta_{TT} = -1$.  The basic variables of
the Ashtekar approach are an inverse densitized triad \E{a}{A}\
and a complex SU(2) connection \A{A}{a}.
\begin {eqnarray}
     \E{a}{A}& =& \sqrt g e^a_A; \\
\label{eq:1.1}
     [\E{a}{A},\A{B}{b}]&=& -\hbar \delta (x-x') \delta ^B_A
                         \delta ^a_b.
\label{eq:1.2}
\end{eqnarray}
$e^A_a$ is the usual 3x3 triad matrix and $e^a_A$ is its inverse.
The local Lorentz indices are vector rather than spinor, and
strictly
speaking the internal symmetry is O(3) rather than SU(2).  After
the  gauge-
fixing allowed by the planar symmetry, the internal symmetry is
O(2) rather than U(1).

\section{Quantization}

     The first part of this section summarizes the main results and
formulas on quantization of plane waves from II, for the
convenience of readers who do not have a copy of II at hand.  The
second part discusses a regularization needed to define
unambiguously the action of one term in
the scalar constraint.

     I follow the quantization procedure used by Husain and Smolin
\cite{HSm}.  These authors
use the planar symmetry  to solve and eliminate four constraints
(the
x and y vector constraint and the X and Y Gauss constraint) and
correspondingly eliminate four pairs of  (\E{a}{A}, \A{A}{a})
components.  The 3x3 \E{a}{A}\
matrix then assumes a block diagonal form, with one 1x1 subblock
occupied by \E{z}{Z}\, plus one 2x2 subblock which contains all
\E{a}{A} with a = x,y and A = X,Y.  The 3x3 matrix of connections
\A{A}{a}\ assumes a similar block diagonal form.  None of the
surviving fields depends on x or y.

     After these simplifications, the total Hamiltonian reduces
to a linear combination of the three surviving constraints,
\begin{eqnarray}
     H_T &=& \Np [\Etwo \epsilon _{AB}\epsilon _{ab}
          (\E{z}{Z})^{-1} \A{A}{a} \A{B}{b}/2
     + \epsilon _{MN} \E{b}{M} {\rm F}^N_{zb}]  \nonumber \\
     & & + iN ^z \E{b}{M} {\rm F}^M_{zb} \nonumber \\
          & & -i{\rm N_G} [\partial _z\E{z}{Z} - \epsilon_{IJ}
                     \E{a}{I} \A{J}{a} ] + S.T.\nonumber \\
     &\equiv & \Np H_S + N^zH_z + {\rm N_G}H_G +
S.T.,\label{eq:2.1}
\end{eqnarray}
where
\begin{equation}
     {\rm F}^N_{zb} = \partial _z\A{N}{b}
                          -\epsilon _{NQ}\A{Z}{z} \A{Q}{b}.
\label{eq:2.2}
\end{equation}
$H_S, H_z$, and $H_G$ are the surviving scalar, vector, and Gauss
constraints.  Strictly speaking these are Hamiltonian densities;
for simplicity I have suppressed
an integration over the z axis.  S.T. denotes ``surface''  terms
(terms evaluated at the two endpoints on the z axis).  The
detailed form of these terms is
worked out in II but will not be needed here.

The theory may be quantized by replacing half of each
conjugate momentum-coordinate pair
by a functional derivative, in the usual way,
\bea
     \A{A}{a} &\rta &\hbar  \delta / \delta  \E{a}{A};\ A = X,Y; a
= x,y;
                              \label{eq:2.2a} \\
     \E{z}{Z} & \rta &- \hbar \delta / \delta  \A{Z}{z},
                              \label{eq:2.2b}
\eea
in order to realize the canonical commutation relations, \EQ{1.2}

     In II I  modified the Husain-Smolin quantization procedure in
certain respects.   Those
authors consider the closed topology (the z axis is topologically
a circle $S_1$) whereas I
consider the open one (the z axis is the real line R).   Persons
familiar with open topology
calculations in three spatial dimensions might suppose that
therefore the constraint \EQ{2.1}
should be integrated over  the entire real axis, $z = -\infty$ to
$z = + \infty$.     In the present
case (effectively one space
dimension rather than three, because of the planar symmetry)  the
space does {\it
not} become flat at z goes to infinity.  Moving the boundary points
to spatial infinity does not
especially simplify matters, and in fact integrals are better
behaved if  all integrations are taken
from $z_0$ to $z_{2n+1}$ (in the notation to be used in
section 3), where the points $z_0$ and
$z_{2n+1}$ are a finite distance from the origin.    The radiation
is assumed to be
wave packets confined to the region $z_0 < z < z_{2n+1}$ inside the
boundaries, so that the
gravitational field is relatively simple (but not flat) at
boundaries.  Again,  the details of the
fields at the
boundary will not be needed here, but  perhaps it is helpful to
point out that the result that the
space does not become flat as z goes to infinity does agree with
one's
intuition from Newtonian gravity, where the potential in one
spatial dimension due to a bounded source grows as z at large z.

     I have modified the Husain-Smolin quantization in another
respect. In
equation~(\ref{eq:2.1}) the Lagrange multiplier \Np\ and the
scalar constraint $H_S $ are rescaled versions of the usual
Lagrange multiplier \ut{N}  and Ashtekar scalar constraint H:
\bea
     \ut{N} H &=& [(\ut{N} \E{z}{Z})][H /\E{z}{Z}]
                         \nonumber \\
           &\equiv& [(\Np)][H_S]
\label{eq:2.3}
\eea
A rescaling of this type, for a one space-dimensional theory, was
first proposed by Teitelboim in
the context of  geometrodynamics \cite{Teinorm}.  This
innocuous-looking rescaling
has profound effects on the
closure of the algebra of constraints.  As shown in II, when the
factors are ordered as in
\EQ{2.1}, with all functional derivatives to the right, then  the
constraint algebra is consistent,
and simultaneously $H_z$ can be interpreted as the  generator of
diffeomorphisms.

      However, since $H_S = H/\E{z}{Z} $, and H is polynomial, the
new scalar constraint $H_S $ is rational.  I view this complication

as a price which must be paid; but it is a small price,
considering what one gets in return.   The constraints close,
which they should as a matter of principle; and the quantum
theory possesses the diffeomorphism invariance  characteristic of
the classical theory .

     Nevertheless, at some point a price must be paid: how
does one define the inverse operator $ (\E{z}{Z})^{-1} $?
Fortunately, for the
wavefunctionals considered in this paper and in II there is a
natural definition of the inverse.  The  variable \A{Z}{z}
conjugate to
\E{z}{Z} occurs only in holonomies
\be
     M(z_{i+1}, z_i) = exp[i\int_{z_i}^{z_{i+1}} S_z \A{Z}{z}(z')
dz'].
\label{eq:2.4}
\ee
$S_z$ is the usual 2j+1 dimensional matrix representation of the
generator for SU(2) rotations
around the Z axis, and there is an explicit factor i because $S_z$
is Hermitean.
From \EQ{2.2b} the action of  \E{z}{Z}
on this function is
especially simple, since it merely
brings down a factor of $-iS_z = -im_z$ where $m_z$ is the
eigenvalue of
$S_z $.   The natural definition of
the inverse $ (\E{z}{Z})^{-1}$
is then
\be
      (\E{z}{Z})^{-1}M =  (-im_z)^{-1}M.
\label{eq:2.4a}
\ee
 This works provided $m_z $ is never
allowed to assume  the value zero.

     So far I have been merely reviewing matters already discussed
in II.   I now consider a
new topic, regularization of the constraints.   As with the
definition of  the $ (\E{z}{Z})^{-1}$
operator just given, the regularization will make the
constraints finite and unambiguous
when the constraints  act on the wavefunctionals considered in this
paper; no claim is made that the proposed
regularization will work in all
circumstances.   I  follow
Husain and Smolin \cite{HSm}, who suggest a simple point splitting
regularization.  This simple technique
works  because the planar case is simpler than
the full three-dimensional
case.  In the full  case, the wavefunctionals typically involve
integrals over loops (Wilson loops)
which are one-dimensional,  while the functional derivatives
analogous to \EQ{2.2a} give rise to
three-dimensional delta functions \cite{Blen,OptrOrd,Bor}.  There
are more delta functions
than integrals, which is a
recipe for divergences.  In the planar case, the wavefunctionals
still involve one-dimensional
integrals ( always over the z axis), but the delta functions from
the functional derivatives are also
one-dimensional.

     Even when the dimensionality of the integrals
matches the dimensionality of the delta functions, it is possible
to get a divergence if the wavefunctional contains a product of two
or more fields evaluated at the same point.  (For example a
wavefunctional of the form $\int dz' [\Etld (z')]^2$ acted upon by
a constraint operator containing a product $[\delta /\delta \Etld
 (z)]^2$ gives rise to an undefined $[\delta (z-z')]^2$ .)  But
this
does not happen in the present case: the wavefunctionals studied
here involve at most one field evaluated at each point.

     In fact only the first term in \EQ{2.1} needs regularization,
and the regularization is needed
to remove an ambiguity rather than a divergence.   This term is the
only one containing three,
rather than one functional derivative.  The wavefunctional is
path-ordered, so that one can talk
meaningfully about ``adjacent'' fields.  When the first term acts
on  three adjacent fields, the
point splitting is needed to define unambiguously which three
fields will be acted upon in which
order.

     I call the first term $H_E$ (E for the \Etwo\ which it
contains) and point split it as
follows:
\be
     H_E =  \Np \Etwo(z) \epsilon _{AB}\epsilon _{ab}
\A{A}{a}(z+\epsilon)
[\E{z}{Z}(z)]^{-1} \A{B}{b}(z - \epsilon)/2 .
 \label{eq:2.5}
\ee
One can now see how the point splitting removes an ambiguity.  It
defines unambiguously which holonomy M the $[\E{z}{Z}]^{-1}$
operator acts on, when     the
wavefunctional contains a chain of holonomies separated by \Etld\
fields.
\be
     \int dz_1 dz_2 \cdots M(z_{i+1},z_i) \Etld (z_i)
M(z_i,z_{i-1}) \Etld (z_{i-1})M \cdots.
\label{eq:2.6}
\ee
The integrals are path ordered,
\be
     \cdots z_{i+1} \geq z_i \geq z_{i-1} \cdots.
\label{eq:2.7}
\ee
When the two operators $A = \delta/\delta \Etld$ in $H_E$ act on
the two \Etld\ in \EQ{2.6}, the
$[\E{z}{Z}]^{-1}$ must act on the holonomy {\it between} the
\Etld, because of  the pattern of
epsilons in $H_E$  .  Another pattern of epsilons would have forced
the $[\E{z}{Z}]^{-1}$
to act on the holonomy to the right or to the left of the two
\Etld.

     I have chosen the pattern in
\EQ{2.5} because of its symmetry and simplicity.  The calculations
of the next section are not
unduly sensitive to this choice: one can still find solutions when
the $[\E{z}{Z}]^{-1}$ acts on
the holonomy to
the left or the right, although they differ slightly from the
solutions when the $[\E{z}{Z}]^{-1}$
acts in the middle.  Note that another symmetric choice yields
nothing new: reversing the sign of
the epsilons in \EQ{2.5} does not change anything, since $H_E$ is
even under interchange of
the A's.   However, I have been unable  to find a solution  using
an average over left, right, and
middle holonomies, rather than the middle holonomy alone.

     I have verified that the pattern of epsilons given in \EQ{2.5}
is preserved by the
constraint algebra, in the following sense.  The $H_E$ term is
contained in the scalar constraint
$H_S$.  Consider a constraint commutator where $H_S$ occurs on both
sides of the equation.
\be
     [H_S, H_z] \propto H_S.
\ee
If the $H_E$ on the left is given the pattern of epsilons shown in
\EQ{2.5} then the $H_E$ on
the right will also follow the same pattern.  (Also, there will be
additional terms of order epsilon
on the right, because the point splitting violates diffeomorphism
invariance; but these terms
disappear in the limit $\epsilon \rta 0$.)  If there is no $H_S$ on
the right,
\[
     [H_S,H_S] \propto  H_z ;
\]
\[
     [H_S, H_G]  =  0,
\]
then the right hand sides will  contain additional terms of order
epsilon, which again
disappear in the limit $\epsilon \rta 0$.

\section{Solutions}

     This section proposes a solution, then verifies that it is
annihilated by the constraints.  In II  the constraints were
expressed in terms of fields which are
eigenstates of the surviving
gauge invariance O(2) or U(1) generated by $H_G$.  That is, I used
basis fields
\begin{equation}
     \E{a}{\pm} = [\E{a}{X} \pm i\E{a}{Y} ]/\sqrt{2} ,
\mbox{a=x,y,}
\label{eq:3.0a}
\end{equation}
rather than the usual \E{a}{X} and \E{a}{Y}; similarly, I used
\A{\pm}{a}.  The solutions of
both II and the present paper are much simpler when expressed in
terms of these basis fields.
Accordingly,  I return to the Hamiltonian, equation~(\ref{eq:2.1})
and
break it up into eigenstates of O(2) by writing out the
components of the Levi-Civita tensor,
\begin{equation}
     \epsilon _{-+} = -\epsilon _{+-} = i,
\label{eq:3.0b}
\end{equation}
while being careful to contract every + index with a - index, for
example $\epsilon _{MN} A^M B^N = \epsilon _{-+} A^{+} B^{-} +
\cdots $.
\begin{eqnarray}
     H_T &=& \frac{1}{2}(\Np-N^z) [-\epsilon _{mn}\E{m}{+} \E{n}{-}
                (\E{z}{Z})^{-1} \epsilon _{cd} \A{+}{c} \A{-}{d}
               -2i\E{b}{-}(\partial _z + i\A{Z}{z}) \A{+}{b}]
                                \nonumber \\
     & & + \frac{1}{2}(\Np + N^z) [-\epsilon _{mn}\E{m}{+} \E{n}{-}
                (\E{z}{Z})^{-1} \epsilon _{cd} \A{+}{c} \A{-}{d}
                    +2i\E{b}{+}(\partial _z -i\A{Z}{z}) \A{-}{b}]
                                   \nonumber \\
      & & -i{\rm N_G} [\partial _z\E{z}{Z} - i(\E{a}{+} \A{-}{a}-
               \E{a}{-} \A{+}{a})] + S.T.
\label{eq:3.0c}
\end{eqnarray}
It is perhaps also worth repeating the quantization equations using
this O(2) eigenbasis, since the
pattern of $\pm$ signs may look unfamiliar.
\begin{eqnarray}
     \A{\pm}{a} &=& \hbar \delta/\delta \E{a}{\mp}; \nonumber \\
     \E{z}{Z}   &=& -\hbar \delta/\delta \A{Z}{z}.
\label{eq:3.0d}
\end{eqnarray}
To understand the pattern of signs, note that the two
dimensional Kronecker delta in \EQ{1.2} has only off-diagonal
elements when
expressed in terms of O(2) eigenstates: $\delta_{\pm \mp} = +1$.)

     The ansatz for the solution contains n factors of
$\E{a}{-}S_{+}$
followed by an equal number of
factors of $\E{a}{+}S_{-}$, where $S{\pm}$ are SU(2) raising and
lowering operators for the
2j + 1 dimensional representation of SU(2).  The \Etld S factors
are separated by the z-axis
holonomies M introduced at \EQ{2.4}.
\bea
          \psi (2n;j) &=& \prod_{i=1}^{2n} \int_{z_0}^{z_{2n+1} }
dz_i \theta (z_{i+1}-z_i)
                       \theta (z_1,z_0) \times  \nonumber \\
          & & \times M(z_{2n+1},z_{2n}) \E{a_{2n}}{+}(z_{2n})S_{-}
     M(z_{2n},z_{2n-1}) \cdots  \E{a_{n+1}}{+}(z_{n+1})S_{-}
M(z_{n+1},z_n )
                                             \times  \nonumber \\
     & & \times  \E{a_n}{-}(z_n)S_{+} M(z_n, z_{n-1} )
\E{a_{n-1}}{-}(z_{n-1})S_{+} \cdots
          \E{a_1}{-}(z_1)S_{+} M(z_1, z_0 )  \times  \nonumber \\
     & & \times M(z_0, z_{2n+1} ) \epsilon_{a_{n+1}a_n}.
\label{eq:3.1}
\eea
 If the holonomy $M(z_{i+1}, z_i )$ is pictured as  a flux line
extending from $z_i$
to $z_{i+1}$, then $\psi$ corresponds to an open-ended flux line
stretching from $z_0$ to
$z_{2n+1}$, with 2n \Etld\ operators inserted along the line.   The
final $M(z_0, z_{2n+1} )$
holonomy  represents a  flux line which loops back from  $z_{2n+1}$
to $z_0$ and  turns
the open-ended flux line into a closed loop.
     The SU(2) structure of \EQ{3.1} supports this interpretation.
The $S_{\pm}$ and M
are  (2j+1) $\times$ (2j+1) dimensional matrices with rows and
columns labeled by the
eigenfunctions $m_z$ of $S_z$.  The $m_z$ subscripts on these
matrices have been suppressed
for simplicity, but
clearly , since there are an equal number of raising and lowering
operators in the chain, the first
and
last holonomies in the chain have the same value of  $m_z$.  This
means that  the return
segment $M(z_0, z_{2n+1} )$   has the right $m_z$ subscripts to
match the
$m_z$ of the flux exiting from
the beginning and end of the open chain.
     Note that it is {\it not} necessary to sum over the $m_z$ (it
is not necessary to take a trace)
because the internal gauge group SU(2) has been fixed to O(2), and
the irreducible representations
of O(2) are all one-dimensional, labeled by their $m_z$ value.
Later I use identities such as
\be
     [S_z,S_{\pm}] = \pm S_{\pm},
\ee
but even here there is no need to sum over the intermediate values
of $m_z$ on the left, since
only one intermendiate value occurs anyway.  If I wished, I could
replace every $S_i$ by the
corresponding matrix element; e.g.\ replace $S_z$ by $m_z$; but
that would be awkward,
especially for the $S_{\pm}$.

     Since $m_z$ is not summed over, the
$m_z$ value at (say) the start of the chain remains a free
parameter for the moment.  Later in
this section we shall see that the
scalar constraint is not satisfied unless the holonomy $M(z_{n+1},
z_n )$ has  $m_z = -1/2$.  This particular
holonomy occurs at the changeover point, where the chain shifts
from $S_{+}$ to $S_{-}$. Note
also the $a_i$ indices on either side of the changeover  point must
be
antisymmetrized by the final epsilon
tensor in \EQ{3.1}.   Again, this is required by  the scalar
constraint.  The free parameters are therefore
n, j, and the $a_i$ away from the changeover point.

     Since every  $m_z$ occurs contracted with another $m_z$, and
every $S_{\pm} $ is contracted with an
\E{a}{\mp}, it should be clear that $\psi $ is annihilated by the
Gauss constraint.  To verify this
explicitly, note from \EQ{3.0c} that this constraint is the sum of
three terms,
\be
     \partial_z \E{z}{Z} + i(\E{a}{-} \A{+}{a} - \E{a}{+}
\A{-}{a}).
\ee
The first term acts only on the holonomies:
\be
     \partial_z \E{z}{Z}(z) M(z_{i+1},z_i) = iS_z[\delta(z_{i+1}-z)
- \delta(z_i - z)].
\label{eq:3.2}
\ee
The remaining two terms act on the $\E{a_i}{\mp}$.  These two
terms, like the first term, also
generate factors of $\pm
i \delta (z_i - z)$, but no factors of $S_z$.  Now group together
all terms in $H_G \psi$
containing a
factor $\delta (z_i - z)$. If $z_i$ is not an endpoint $z_0 $ or
$z_{2n+1}$, there will be three such
terms, one from \E{a_i}{\pm}, and two from the holonomies on either
side of  \E{a_i}{\pm}:
\be
     i \delta (z_i - z) M(z_{i+1},z_i)\{ \pm S_{\mp} +[S_z,S_{\mp}]
\} \E{a_i}{\pm}
          M(z_i,z_{i-1}) \cdots.
\ee
From the commutation relations obeyed by the $S_i$, the curly
bracket vanishes.  If $z_i$ is an
endpoint $z_0 $ or $z_{2n+1}$,
there will be no contributions from \Etld\ fields, only
contributions  from the three holonomies in
$\psi$ which depend on the endpoints.  But these holonomies can be
combined so that all dependence
on the endpoints disappears:
\be
     M(z_{2n+1},z_{2n}) \cdots  M(z_1, z_0 )M(z_0, z_{2n+1} ) =
M(z_{2n},z_1).
\ee
Hence there is no contribution from the endpoints either, and the
Gauss constraint annihilates $\psi$.

     Now consider the diffeomorphism constraint $H_z$.  The proof
that this annihilates $\psi$
proceeds along the same lines as the corresponding proof for the
solutions presented in II, since again
$\psi $  is a product of diffeomorphism-invariant integrals of the
form $\int dz_i$(scalar density \Etld).   I mention only the one
point
of the proof which is not
straightforward.  Since the range of integration is $z_0$ to
$z_{2n+1}$, rather than the whole real axis,
the points   $z_0$ and $z_{2n+1}$ are singled out.   This seems to
violate diffeomorphism
invariance.  However, the precise statement of this invariance
requires that the wavefunctional be
annihilated by the {\it smeared} constraint $\int \delta N_z H_z
dz$, where $\delta N_z $ is an
infinitesimal change in the shift vector.   Away from boundaries
$\delta N_z $ is arbitrary, but at
boundaries it must vanish, in order to preserve the boundary
conditions  $N_z = 0$.    This vanishing is enough
to guarantee the vanishing of any contributions to $\int \delta N_z
H_z dz \psi$ from boundary points, and
the proof of diffeomorphism invariance is not affected by the
finite range of integration.

     Only the scalar constraint remains to be considered.  From
\EQ{3.0c} this
can be written as a sum of three terms.
\be
     H_S = H_{+} + H_{-} + H_E,
\label{eq:3.3}
\ee
where
\be
     H_{\pm} = \pm i \Np \E{a}{\pm} ( \partial _z  \mp i \A{Z}{z})
\A{\mp}{a};
\label{eq:3.4}
\ee
\be
     H_E = \Np [\Etwo (z) \epsilon _{AB} \epsilon _{ab}
\A{A}{a}(z+\epsilon)
                     (\E{z}{Z}(z))^{-1}  \A{B}{ b}(z-\epsilon)/2 .
\label{eq:3.5}
\ee
It is understood that the $\A{A}{a}$ and $\E{z}{Z}$ operators in
these equations are to be replaced by the functional derivatives
given at equations~(\ref{eq:3.0d}).  Consider the action of
$H_{\pm}$ first.  This
operator acts on the factors of $\E{a_i}{\pm}$ in $\psi$ but does
not remove the \Etld;
instead, it removes one of the adjacent holonomies.
\bea
H_{\pm}\psi &=& H_{\pm}[ \cdots \theta (z_{i+1}-z_i)
                    \theta (z_i-z_{i-1}) M(z_{i+1},z_i )
               \E{a_i}{\pm}(z_i)S_{\mp} M(z_i, z_{i-1} ) \cdots ]
                                   \nonumber \\
     &=&  \sum_i[\cdots \theta (z_{i+1}-z_i)  \theta (z_i-z_{i-1})
               M(z_{i+1},z_i ) (\pm i \Np \E{a_i}{\pm}(z)) \times
                    \nonumber \\
     & & \times ( \partial _z  \mp i \A{Z}{z}) \delta (z-z_i)
               S_{\mp} M(z_i, z_{i-1} ) \cdots] .
\label{eq:3.6}
\eea
Now change the $\partial_z$ on the delta function to a
$-\partial_{z_i}$ and integrate by parts
on $z_i$.  The surface terms at $z = z_0$, $ z_{2n+1}$ again vanish
because the smearing function
\Np\ is actually a $\delta \Np$ which vanishes at boundaries.  The
$\partial_{z_i}$ acts on
the M  and the $\theta$ factors in \EQ{3.6}.  Since The action on
the M's
brings down factors of  \A{Z}{z}, group these terms  with the term
in \EQ{3.6} which already
contains an \A{Z}{z}:
\bea
     ( \partial_{z_i} \mp i \A{Z}{z}) [M(z_{i+1},z_i )  (\pm i)
S_{\mp} M(z_i, z_{i-1} )]           &\propto &  i \A{Z}{z}) [-iS_z
S_{\mp}+ S_{\mp}iS_z \mp S_{\mp}]
                                   \nonumber \\
          &=& 0.
\eea
Thus the  \A{Z}{z} term in \EQ{3.4} and \EQ{3.6} cancels terms
where derivatives act on factors of M,
leaving the terms where derivatives act on the $\theta$ functions.
\bea
H_{\pm}\psi &=&\pm i \Np \sum_i [ \cdots \delta (z-z_i)
               (\theta (z_{i+1}-z_i)  \delta (z_i-z_{i-1}) -
               \delta(z_{i+1}-z_i) \theta (z_i-z_{i-1})) \times
                              \nonumber \\
          & & \times M(z_{i+1},z_i ) \E{a_i}{\pm}(z) S_{\mp}
                    M(z_i, z_{i-1} ) \cdots ].
\label{eq:3.7}
\eea
The two delta functions in each term reduce one of the M's to
unity, in effect removing it
completely.  As stated previously, the $H_{\pm}$ terms do not
remove an \Etld, but do remove a
neighboring holonomy.

     Now consider the $H_E$ term in the scalar constraint.  From
\EQ{3.5}, this removes two
\Etld\ and replaces them by an \Etwo.  The two \Etld\ which are
deleted must be adjacent along
the chain; otherwise there will be two delta functions
$\delta(z-z_i) \delta(z-z_j)$ which will
squeeze the range of integration of a  $dz_k$ to zero, where $z_k$
lies between $z_i$ and $z_j$.
Because of the $\epsilon_{AB} $ in \EQ{3.5}, one of the adjacent
\Etld\ must be an \E{a}{+},
and the other must be an \E{a}{-}.  Therefore $H_E$ acts only at
the changeover point, where
the chain shifts from $S_{+}$ to $S_{-}$.
\begin{eqnarray*}
     H_E \psi & = & \cdots  \Np [\Etwo (z) \epsilon _{+-}
               \epsilon _{a_{n+1}a_n} \delta (z+\epsilon-z_{n+1})
               \delta (z-\epsilon -z_n)/2 ] \times \\
          & & \times (\E{z}{Z}(z))^{-1} S_{-} M(z_{n+1},z_n ) S_{+}
                     \cdots  \epsilon _{a_{n+1}a_n} \\
          &=&  \cdots  \Np [\Etwo (z) \epsilon _{+-}(
          \epsilon _{a_{n+1}a_n})^2 \delta (z+\epsilon-z_{n+1})
               \delta (z-\epsilon -z_n)/2 ] \times \\
          & & \times  S_{-} (-im_z)^{-1}M(z_{n+1},z_n ) S_{+}
\cdots
\end{eqnarray*}
I have  used \EQ{2.4a} for the
action of $\E{z}{Z}(z)^{-1}$ on M. Now take the limit $\epsilon
\rta 0$.
\bea
    H_E \psi & \rta & \cdots  \Np [\Etwo (z) (-i) (+2)
               \delta (z-z_{n+1})  \delta (z- -z_n)/2 ] \times
                              \nonumber \\
          & & \times  S_{-} (-im_z)^{-1}(1) S_{+} \cdots
                              \nonumber \\
          &=& \cdots  \Np [\Etwo (z)  \delta (z-z_{n+1})
               \delta (z- -z_n) ] S_{-} (m_z)^{-1} S_{+} \cdots
\label{eq:3.8}
\eea
As predicted in the last section, the $z \pm \epsilon$ point
splitting is not needed to regulate
$(\delta  (z- -z_n))^2$ divergences; there are none.  Rather, it is
needed to define unambiguously
which M the $\E{z}{Z}(z))^{-1}$ is to act on: it acts on the M
between the two \Etld\ grasped by
the A's in $H_E$.

     The action of $H_{\pm}$ and $H_E$ on $\psi$ has now been
described; it is time to put the pieces together and describe the
action of the entire scalar constraint on $\psi$.  Each term in
$H_S \psi$ contains a double delta function $\delta (z-z_{i+1})
\delta (z-z_i)$; cf. \EQ{3.7} and \EQ{3.8}.  It is therefore
natural to group together terms having the same double delta
function.  Terms where $z_{i+1}$ or $z_i$ is a boundary point can
be dropped, because \Np(z) (or more precisely, $\delta \Np(z)$ )
vanishes at boundary points.
For the moment ignore the  $\delta (z-z_{n+1}) \delta (z- z_n)$
terms at the crossover point, since these are the only terms
to receive a contribution from the $H_E$ piece of $H_S$.  $H_S
\psi$ contains only two terms proportional to $\delta (z-z_{i+1})
\delta (z- z_i)$, $i \neq n$.  One of the two comes from the term
shown explicitly in \EQ{3.7}.  This term is the ith term in a sum
over 2n terms.  The other $\delta (z-z_{i+1}) \delta (z- z_i)$
term comes from the (i+1)st term in the same sum.  The two terms
have opposite sign and so cancel.

     Now consider the $\delta (z-z_{n+1}) \delta (z- z_n)$ terms,
those at the crossover point.  Since the crossover point is flanked
by both an \E{a_{n+1}}{+} (at $z=z_{n+1}$)and an \E{a_n}{-} (at
$z=z_n$), there will be contributions from both $H_{+}$ and
$H_{-}$.   For the $H_{+}$ contribution, set i = n+1 in \EQ{3.7};
for the $H_{-}$ contribution, set i = n.  These two contributions
have the same sign and do not cancel.
\bea
     (H_{+} + H_{-})\psi & = & \cdots \Np \delta (z-z_{n+1})
               \delta (z- z_n) 2i \E{a_{n+1}}{+}(z) S_{-}
               \E{a_n}{-}(z) S_{+} \cdots \epsilon_{a_{n+1}a_n}
                         \nonumber \\
          &=& \cdots \delta (z-z_{n+1})
               \delta (z- z_n) 2 \Etwo(z) S_{-} S_{+} \cdots.
\label{eq:3.9}
\eea
This will be canceled by the $H_E$ contribution, \EQ{3.8}, provided
one chooses
\be
     m_z = -1/2.
\label{eq:3.10}
\ee
This must be the value of $m_z$ at the changeover point.  At the
$z=z_0$
end of the chain, then, $m_z = -1/2 - n$.  The n $S_{+}$ operators
raise this to a maximum of -1/2 at the changeover point; then the
n $S_{-}$ operators lower this back to $m_z = -1/2 - n$ at the $z
= z_{2n+1}$ end of the chain.  At no point does $m_z$ pass through
the forbidden value of zero.   This completes the proof that
$\psi$, \EQ{3.1}, is annihilated by all the constraints.

     From this solution one can generate others.  E.\ g.\, by
interchanging plus and minus SU(2) subscripts in \EQ{3.1}, one
generates the "complex conjugate" wavefunctional, which is also a
solution with all $m_z$ positive and \EQ{3.10} replaced by $m_z =
+1/2$.  Also, it is possible to relax the requirement that all
$S{\pm}$ and M have the same j; see the concluding section of II.

\section{Discussion and Directions for Further Work}

           What is the physical interpretation of the solutions of
section 3?
 In II I argued that, since $H_S$ generates
time translations and $H_z$ generates space translations, it is
reasonable to assume that the operators $H_S \pm H_z$ represent
displacements
along the light cone directions $t' \pm z'$ intrinsic to the plane
wave metric.  (In technical terms  the metric considered here
posseses two hypersurface
orthogonal null vectors; and the corresponding hypersurfaces can be
parameterized by $t' \pm z'$\ = constant, where the coordinates
(t',z') are related to (t,z) by an
appropriate gauge transformation.)  As a check, one can
show that the commutator $[H_S + H_z, H_S - H_z]$ vanishes, as it
should if it is the commutator of two independent translation
generators.  Note $(H_S \pm H_z)/2$ is the operator multiplying
$\Np \pm N^z$ in \EQ{3.0c},
and this operator is linear in \A{\mp}{a} = $\hbar \delta/\delta
\E{a}{\pm}$.   The solutions in II
depend on either \E{a_i}{-} or \E{a_i}{+} fields, but not both.
This implies that  the solutions  obey either $(H_S +H_z)\psi = 0$
or $(H_S - H_z)\psi = 0$,
which suggests that they depend on either $ t'-z'$ or $t'+z'$, but
not both.  If the solutions of II
represent radiation, then that radiation is unidirectional,
traveling in a single direction along the
z axis.  The solutions constructed in the present paper contain
both \E{a_i}{-} and
\E{a_i}{+} fields.  By the same argument, if the solutions of this
paper represent radiation, it is
not unidirectional.  These solutions correspond to scattering.

     The argument just outlined is very limited.  It does reveal
the direction of motion of the
field.  By itself, however, it cannot establish that the field is
radiative, much less give details of
the polarization and amplitude of the radiation.  One can ask to
what extent it is possible to go
farther and construct operators which explore these details.   In
the 1970's, a great deal of work
was done on characterization of {\it classical} radiative plane
wave solutions, and most of this
work  is readily translated into the Ashtekar language.  However,
it is one thing to
characterize classical radiation, and another to translate a
classical criterion into a quantum
criterion.

     It is not hard to see what issues can arise.  For simplicity,
consider the corresponding
problem in electrodynamics: given a radiation criterion valid for
classical Maxwell
electrodynamics, translate it into the corresponding criterion for
QED.   Consider, for example ,
the classical criterion that a plane wave traveling in the z
direction has only a single polarization:
\be
     (E_x + iB_y)_{cl} = 0.
\label{eq:4.1}
\ee
The corresponding quantum criterion is not
\be
     (E_x + iB_y)   \psi = 0,
\label{eq:4.2}
\ee
where $\psi$ is now a state in the Hilbert space of the operators
E and B.  To obtain the quantum
analog of the classical equation \EQ{4.1}, one must pass to large
occupation numbers {\it and}
use coherent states:
\be
     <coh \mid (E_x + iB_y) \mid coh > = 0.
\label{eq:4.3}
\ee
\EQ{4.2} is far too strong: the vacuum state with respect to a
given polarization need not be
annihilated by the corresponding operator.    \EQ{4.3}, however, is

too special, since it works
for coherent states but not for eigenstates of the number operator.

If one has the full machinery
of QED, of course the correct procedure is to break up E and B into
creation and annihilation
operators and demand that \^{a}$\psi = 0$,  \^{a} an annihilation
operator for  the relevant
polarization.   In a generally covariant theory, creation and
annihilation operators are not
automatically available, since normally  a preferred time
coordinate is required for their
definition, e.\ g.\ the time associated with the timelike Killing
vector in a static spacetime.   Here
one has the preferred coordinates $t'\pm z'$, but in these
coordinates the metric is conformally flat rather than static.   On
the other hand, unidirectional plane waves are known to obey a
superposition principle \cite{Bon,Aich}, which suggests that the
unidirectional case may possess particle-like modes.  Even if  no
quantum criteria are forthcoming, the classical criteria by
themselves are quite useful and illuminating, and  I
plan a paper which will (at a minimum) translate the classical
criteria for plane radiation into the
Ashtekar language.

\end{document}